\documentstyle[multicol,prl,epsf,aps]{revtex}   

\begin{document} \draft

\title{Effect of order-parameter fluctuations on
the Halperin-Lubenski-Ma first-order transition in superconductors
and liquid crystals}

\author{Igor F. Herbut, Anand Yethiraj, and John Bechhoefer}

\address{Department of Physics, Simon Fraser University, 
Burnaby, British Columbia, \\
 Canada V5A 1S6\\}\maketitle

\begin{abstract}
We show that order-parameter fluctuations in a good type-I
superconductor or a liquid crystal always increase the
size of the first-order transition. This behavior is eventually changed when
the system crosses over to inverted-XY critical behavior, with 
the size of the first-order transition vanishing as a power law with a
crossover exponent. We find good agreement between our theory
and a recent experiment on the nematic-smectic-A first-order transition
in 8CB-10CB mixtures of liquid crystals.

\end{abstract}
\vspace{10pt}
\begin{multicols}{2}
\narrowtext

\begin{figure}
\centerline{\epsfxsize=8cm \epsfbox{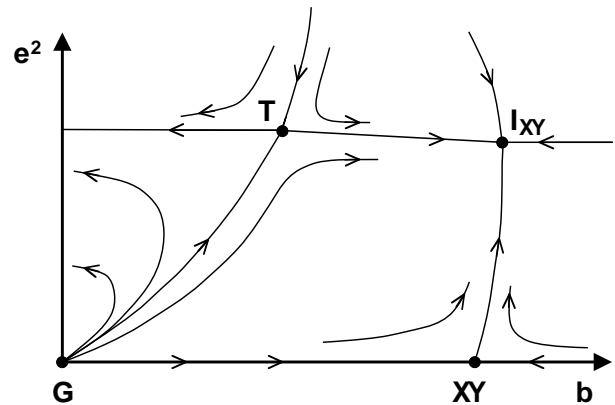}} 
\vspace*{3em} 
\caption{Schematic flow diagram of the quartic coupling $b$ and the 
charge $e^2$ (see Eq. \ref{eq:hamiltonian}) at the critical surface 
$T = T^* $.  Note the 
Gaussian ($G$), XY, inverted-XY ($I_{XY}$), and the tricritical ($T$) 
fixed points.  Good type-I materials ($\kappa \ll 1$) lie far to the 
left of the separatrix that connects the Gaussian and the tricritical 
fixed points.}
\label{flow-diagram}
\end{figure}

  More than 25 years ago, Halperin, Lubensky and Ma (HLM) \cite{halperin}
and Coleman and Weinberg  \cite{coleman} demonstrated that when a scalar 
field is coupled to a gauge field, the fluctuations of the gauge field 
can change the nature of the phase transition in the theory
from continuous to first order. The coupling of a scalar field 
(or an order parameter (OP)) to a massless gauge field arises often
in physics: the Meissner transition in superconductors \cite{ginzburg},
the Higgs mechanism in particle physics
\cite{higgs},
and the nematic-smectic-A transition in liquid crystals \cite{degennes} 
are among the best-known examples \cite{other1}, \cite{other2}.
The analysis of HLM initially centered
on the fluctuations of the gauge field and neglected the
OP fluctuations, which is justifiable for good type-I
superconductors (with Ginzburg-Landau parameter $\kappa \ll 1$),
where the size (to be precisely defined shortly)
of the first-order transition is larger. This approximation, however,
inevitably breaks down for strong type-II materials ($\kappa \gg 1$),
where neglecting the OP fluctuations would yield a first-order transition
well into the critical region.  
Close to four dimensions, the effect of the OP fluctuations can be
studied using the Wilson-Fisher renormalization group \cite{halperin}, which,
however, in this case leads only to ``run-away" flows and no stable critical
points. This is usually interpreted as a sign of a 
first-order transition \cite{halperin}, \cite{lawrie}. Today, 
based on accumulated analytical and numerical evidence,
it is generally believed that the transition for $\kappa \gg 1$
is again second order, in the so-called inverted-XY universality class
\cite{dasgupta,bartholomew,herbut,kleinert,herbut1,bergerhoff,teitel,sudbo}.
The corresponding  topology of the flow of the coupling constants 
under scaling transformation is depicted in Fig. \ref{flow-diagram}. 
It may thus seem natural to  expect that the OP fluctuations should 
decrease the size of the first-order transition, finally reducing it 
to zero at the crossover to inverted-XY critical behavior.

The fluctuation effects in question are unfortunately too fine to be 
observable even in high-$T_c$ superconductors on account of the smallness 
of the fine-structure constant \cite{fisher,kamal}.
But in liquid crystals, the coupling of the smectic OP to the
director fluctuations is stronger, and the fluctuations at the 
nematic-smectic-A transition become an issue of central importance.
A recent experiment on two-component liquid-crystal mixtures 
\cite{yethiraj} found a surprising result:  the size of the first-order
transition is {\it larger} than the prediction of the HLM theory.
Motivated by this result, in this Letter we consider theoretically  
the effect of OP fluctuations on the first-order
transition in a type-I material. We show quite generally that for a small 
enough Ginzburg-Landau parameter $\kappa$, the size of the first-order 
transition is indeed always {\it larger} than the HLM result.
Crudely, the reason is that for a good type-I material, the effective 
Ginzburg-Landau parameter $\kappa$ always decreases at large scales, 
making the material only more type I and the transition more strongly  
first order. The type-I region is defined here as left of the separatrix 
in Fig. \ref{flow-diagram}, where the flow is qualitatively the 
same as that of HLM near four dimensions. This behavior is eventually 
changed as the separatrix between type-I (first-order) and type-II 
(inverted-XY) regimes is approached, where the 
size of the transition goes to zero as a power law,
with a crossover exponent that characterizes the tricritical point.
Using the simple one-loop recursion relations for the coupling constants
in an isotropic version of the 

\noindent
Ginzburg-Landau-Wilson action, we show that the experimental data of 
Yethiraj and Bechhoefer are fit rather well by our theory.

We are interested in a general phase transition described by the
three-dimensional Ginzburg-Landau-Wilson theory for a fluctuating
complex OP  $\Psi(\vec{r})$ minimally coupled to a fluctuating
vector potential (Higgs scalar electrodynamics): 
\begin{eqnarray}
     H = \int d^3 \vec{r} [ |(\nabla
 - i e \vec{A}(\vec{r}))\Psi(\vec{r})|^2 +
     \frac{a(T)}{2}|\Psi(\vec{r})|^2 \nonumber \\
     + \frac{b}{2} |\Psi(\vec{r})|^4 + \frac{c}{6} |\Psi(\vec{r})|^6 +
     \frac{1}{2} (\nabla \times \vec{A} (\vec{r}))^2 ] ,
\label{eq:hamiltonian} 
\end{eqnarray}
where $a(T) = \alpha (T-T^*)/T^* $, and we assume the
superconducting gauge,
$\nabla\cdot\vec{A} = 0$. For simplicity, we neglect the
anisotropy inherent to liquid crystals but retain the sixth-order
term in the Ginzburg-Landau expansion,
which will be needed later for comparison with the experiment. Fields and
lengths have been chosen so that the number of couplings in the theory
is minimal. 
HLM showed that neglecting the fluctuations of the OP and 
integrating out the vector field
$\vec{A}$ around the uniform OP configuration $\Psi_0$,
to the lowest order in charge $e$ one obtains the corrected 
mean-field Ginzburg-Landau free-energy per unit volume:
\begin{equation}
F =\frac{a(T)}{2} \Psi_0 ^2 - \frac{e^3 \sqrt{2}}{3\pi}\Psi_0 ^3 +
     \frac{b}{2} \Psi_0 ^4 + \frac{c}{6} \Psi_0 ^6 ,
\label{eq:free-energy} 
\end{equation}
in units where $k_B T^*  =1$. The negative cubic term in the 
free-energy implies a first-order transition. A useful measure of 
the {\it size} of the first-order transition is the parameter 
$t= (T_c - T^* )/T^* $, where $T_c $ is the first-order
transition temperature. To keep the algebra simple, we will assume  
$b>0$ and temporarily set $c=0$ in Eq. \ref{eq:free-energy}, 
and turn $c$ back to a finite value only later when we compare
our results with the experiment. From Eq. \ref{eq:free-energy},
the size of the transition is then
\begin{equation}
    \alpha t_{HLM} = \frac{e^4}{(6\pi)^2} F(\kappa) ,
\label{eq:alpha-hlm}
\end{equation}
where $\kappa^2 = b/2e^2$ is the dimensionless Ginzburg-Landau parameter,
and $F(\kappa) = 1/\kappa^2$ \cite{note1}. For $\kappa\ll 1$, the first-order
transition thus occurs at higher temperatures, where the OP fluctuations
may indeed be neglected.  The HLM result becomes asymptotically correct
in this limit.

To include the OP fluctuations, we assume that the first-order
transition in the theory (Eq. \ref{eq:hamiltonian}) 
with the coupling constants $e$ and $b$ occurs at some $t\neq t_{HLM}$. 
The renormalizability of the theory (Eq. \ref{eq:hamiltonian})
implies that rescaling  the cutoff $\Lambda\rightarrow
\Lambda/s$ by an arbitrary factor $s>1$ is equivalent to changing 
the coupling constants into $e(s)$ and  $b(s)$ \cite{com},
with the first-order
transition for these renormalized couplings occurring at some new 
temperature $t_0$.  The variation of the renormalized
couplings $e(s)$, $b(s)$, and the temperature $t(s)$ with $s$ is described 
by differential recursion relations of the form 
\begin{equation}
      \frac{d \lambda(s)}{d\ln(s)}= \beta_{\lambda} 
      \left[ e^2(s), b(s), t(s) \right] , 
\label{eq:flow-eqs}
\end{equation}
where $\lambda = \{e^2, b, t\}$. If $t_0$ is large enough, the OP 
fluctuations at the first-order transition in the rescaled theory indeed 
become negligible, and $t_0$ may be {\it approximated} by the
mean-field (HLM) expression:
\begin{equation}
     \alpha t_0 \approx \frac{e^4(s)}{(6\pi)^2} F(\kappa(s)) .
\label{eq:alpha-t0}
\end{equation}
Together with Eqs. \ref{eq:alpha-hlm} and \ref{eq:flow-eqs},
and the boundary conditions $t(1)=t$ and $t(s)=t_0$, Eq. 
\ref{eq:alpha-t0} determines implicitly the actual size of the transition $t$.
Clearly, the above idea is quite general and applicable to other
weakly first-order transitions. 
First, let us demonstrate that neglecting the interactions between
fluctuations
by renormalizing the coupling constants only according to dimensional 
analysis gives just the HLM result.
Power counting in Eq. \ref{eq:hamiltonian}
implies that $t(s) = t s^2$, $e^2 (s) =
e^2 s$, $b(s) = b s $.  Thus, $\kappa(s)= \kappa$, and dividing
Eqs. \ref{eq:alpha-hlm} and \ref{eq:alpha-t0} gives $t=t_{HLM}$.  
More generally, for small $e^2$ and $b$, and for $t\ll e^2, b$,
the renormalized couplings obey the differential
equations (Eq. \ref{eq:flow-eqs})  with \cite{halperin,herbut}:
\begin{equation}
     \beta_t = t( 2+ u e^2 - v b +{\cal O}(e^4, b^2, b e^2) ) ,
\label{eq:beta-t} 
\end{equation}
\begin{equation}
     \beta_e = e^2 - x e^4 +{\cal O}(e^6) ,
\label{eq:beta-e} 
\end{equation}
\begin{equation}
     \beta_b = b-y b^2 + z b e^2 -w e^4 + {\cal O}(b^3, e^6, b e^4, b^2 e^2) .
\label{eq:beta-b} 
\end{equation}
The signs in the above equations are chosen 
so that the numerical coefficients $u$, $v$, $x$, $y$, $z$ and $w$
are positive. Their values in principle are non-universal and
weakly dependent on renormalization procedure. The differential
recursion relation for the Ginzburg-Landau parameter
$\kappa$, for $\kappa \ll 1 $, may be easily obtained from Eqs. 
\ref{eq:beta-e} and \ref{eq:beta-b}: 
\begin{equation}
     \frac{d \kappa^2}{d\ln(s)} = -\frac{w}{2} e^2 +{\cal O}(e^4, e^2 \kappa^2).
\end{equation}
Note that the right-hand side in the last equation is negative, and
$\kappa$, if initially small enough, always decreases under renormalization.
This is a consequence of the $e^4$ term in Eq. \ref{eq:beta-b}, 
which generates a {\it negative} quartic coupling and tends 
to drive the transition first order.
The same term is responsible for the runaway flows near four dimensions
in the original HLM analysis. For small charge $e$, one may neglect
the nonlinear terms and solve Eqs. \ref{eq:beta-t} and \ref{eq:beta-e} 
to find, for small $\kappa$,
\begin{equation}
     \frac{t}{t_{HLM}}=\frac{\kappa^2}{\kappa^2 (\sqrt{t_0/ t})} .
\end{equation}
Since $\kappa(s)<\kappa$, the solution to the above equation always 
gives $t>t_{HLM}$.  In sum, while near the Gaussian fixed point $t$ and 
$e^2$ approximately scale according to dimensional analysis,
a small Ginzburg-Landau parameter
$\kappa$ acquires a small negative dimension and renormalizes downwards,
thus increasing the size of the first-order transition. 

Assuming the general topology of the flow diagram as shown in Fig. 
\ref{flow-diagram} the size of the first-order transition
will go to zero as the coupling constants approach the separatrix
between type-I and type-II regimes.
If the separatrix at the critical surface $t=0$ 
lies at some $b_c(e)$, close to the separatrix one finds
\begin{equation}
     t \propto \left[ b_c (e) -b\right] ^{1/\phi} ,
\label{eq:tricritical}
\end{equation}
where $\phi = r \nu$ is the {\it crossover} exponent, with
$\nu^{-1} = \partial \beta_t /\partial t$, taken at
the {\it tricritical} fixed point. Here, $r$ is the (positive)
scaling dimension of the
second relevant scaling variable at the tricritical point, which 
in general is a linear combination of $e^2$, $b$, and $t$. Near
four dimensions, our result agrees with that of HLM:   
the transition is then always first order, unless $e=0$. The role 
of the tricritical point near four dimensions is
played by the $XY$ fixed point, which has two relevant
directions, $t$ and $e^2$, with $\nu = \nu_{xy}$ and
$r =\varepsilon$, where $\varepsilon = 4-d$. Instead of 
Eq. \ref{eq:tricritical}, one then
has $t\propto e^{2/(\varepsilon \nu_{xy}) }$, which is just Eq. 17
in \cite{halperin}. 

\begin{figure}
\centerline{\epsfxsize=10cm \epsfbox{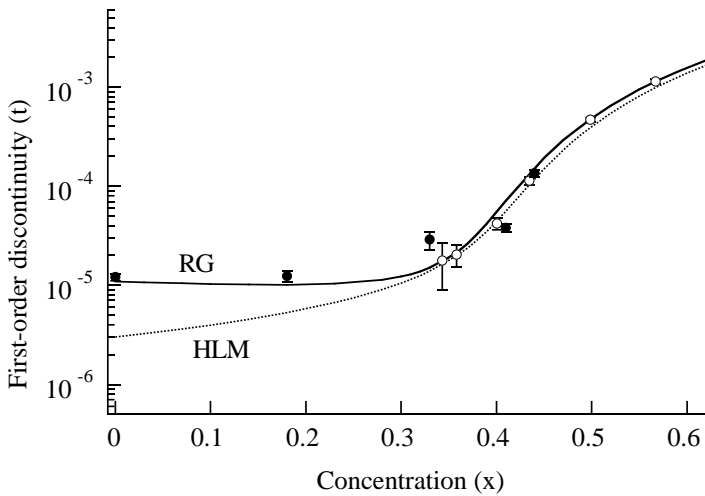}} 
\vspace*{3em} 
\caption{Size of the first-order NA transition ($t$) as a function of 10CB 
concentration ($x$) in the 8CB-10CB system.  
Comparison between the RG result (full line), HLM (dashed line), 
and experiment. Filled circles are data from Yethiraj and Bechhoefer, while 
hollow circles are latent-heat data from Marynissen {\it et al.},  
converted into equivalent size $t$.}
\label{fig:t0-data}
\end{figure}

By tuning the concentration $x$ of 10CB liquid crystals in 8CB, 
one can vary the temperature range of the nematic phase, and, in effect,
tune the parameter $b$ in Eq. \ref{flow-diagram}. 
For larger concentrations, $b$ is negative and the transition is 
more strongly first order. The latent-heat data for $x>0.42$
are well-fit by mean-field theory (Eq. \ref{eq:free-energy}) 
with the parameters $c=1$, $\alpha= 3.35$, $b=b_0 (x^* -x)$,
with $b_0 = 0.395$ and $x^* = 0.42$, and $e^2 = 0.0421$
\cite{marynissen,anisimov,lelidis}. For $x<x^*$,
the quartic term in Eq. (\ref{eq:free-energy})
becomes positive, but the transition remains first order, 
in agreement with the HLM theory. 
At smaller concentrations, $t$ continues to decrease, as expected, 
but there is a clear deviation from the HLM result \cite{yethiraj}. 
To attempt to fit the data in the whole concentration range $0<x<0.65 $,
we take the above parameters to set the initial values of the couplings,
and then evolve $t$, $e$, and $b$ according to Eqs. 
\ref{eq:beta-t}-\ref{eq:beta-b}, with
$u=v=1/4$, $x=1/16$, $y= (2\sqrt{2} +1)/8$, $z=1/2$, and $ w=1/(2\sqrt{2})$
\cite{herbut}. We neglect the change of $c$ under scaling,
since the results are quite insensitive to its precise value.
The only free parameter left is the final value of the 
temperature $t_0$ at which it becomes safe to neglect the 
OP fluctuations. We chose $t_0 = 0.01$, an
order of magnitude larger than the largest measured $t$ in the
experiment. The quality of the fit turns out not to be
critically dependent on this choice. The fit to our theory and the 
comparison with HLM is shown on Fig. \ref{fig:t0-data}. 
Note that for $x>0.35$, there is very little change from the HLM result 
\cite{merging},
but as $b$ increases at smaller concentrations, deviations from  
HLM become significant, in agreement with the experiment \cite{com1}. One 
expects that if one could increase $b$ further, eventually $t$ would
go to zero in accord with Eq. (\ref{eq:tricritical}).
The full non-perturbative structure of the flow diagram in Fig. 
\ref{flow-diagram}, with the tricritical and the inverted-XY fixed points, 
however, is beyond the simple one-loop $\beta$-functions we 
used \cite{comment}. The data suggest that, at small charge, the 
tricritical point is likely to lie at some $\kappa_c > \sqrt{2}$, 
which corresponds approximately to the smallest 
concentration used in the experiment ($x=0$).

Although anisotropy is known to be important in liquid crystals
\cite{degennes}, its inclusion, besides introducing two new couplings,
does not essentially change the structure of Eqs. \ref{eq:beta-t}-
\ref{eq:beta-b} \cite{chen}. In particular, the sign of the
$e^4$ term in Eq. \ref{eq:beta-b} stays negative, and our main point
that the size of the transition in good type-I materials
is larger than the mean-field prediction remains valid. Some small 
quantitative differences from our result would be expected, however.

In conclusion, we have shown that the naive expectation that 
order-parameter fluctuations would always decrease the strength of the 
first-order transition predicted by Halperin, Lubensky, and Ma is 
incorrect for good type-I materials.
While the transition strength does decrease monotonically as the
material becomes more type II, it becomes smaller than 
the HLM prediction only near the tricritical point that separates
type-I and type-II regimes. By using one-loop renormalization group to
calculate the evolution of Landau coefficients in the free energy, we 
can account for the deviations from HLM predictions that were observed 
in a recent experiment.

Our arguments also lead one to expect a non-trivial exponent near the 
tricritical point where the transition eventually should become second 
order.  One would expect this point to be accessible experimentally.  
For example, one can view the variation of 
concentration $x$ in the 8CB-10CB experiments as a way of continuously 
adjusting the effective molecular length (between 8 and 10 CH${}_2$ 
lengths).  Mixing 8CB with a shorter molecule (for example, 6CB 
\cite{even-odd}) might then give access to the tricritical point and 
allow at least a crude measurement of the crossover exponent $\phi$.

This work has been supported by NSERC (Canada). IFH was also
supported by an award from Research Corporation.



\end{multicols}

\begin{thebibliography}{99}
\bibitem{com} We assume that all other couplings are irrelevant.
\bibitem{com1} There is actually a slight increase of the calculated
$t$ at the lowest concentrations, which is likely an artifact of the 
first-order perturbation theory for the $\beta$-functions.
\bibitem{comment} See \cite{herbut} for a modified one-loop
calculation that yields a flow diagram with correct topology.
\bibitem{even-odd}  Because of well-known even-odd variations, the 
properties of the even-numbered chains are more similar to each other 
than are neighboring even- and odd-length molecules.
\end{thebibliography}
\end{document}